# An Improved Method of Lifetime Measurement of Nuclei in Radioactive Decay Chain

J. M. Puzović[a], D. Manić[a], and L. J. Nađđerđ[b]

[a]*Faculty of Physics, University of Belgrade, Studenstki trg 12, 11000 Belgrade, Serbia*
[b]*Institute of Nuclear Sciences "Vinča", University of Belgrade, Mike Alasa 12, 11307 Vinca-Belgrade, Serbia*

We present an improved statistical method for the calculation of mean lifetime of nuclei in a decay chain with an uncertain relation between mother and daughter nuclei. The method is based on the formation of time distribution of intervals between mother and daughter nuclei, without trying to set the exact mother-daughter nuclei relationship. If there is a coincidence of mother and daughter nuclei decays, the sum of these distributions has flat term on which an exponential term is superimposed. Parameters of this exponential function allow lifetime of daughter nucleus to be extracted. The method is tested on Monte Carlo simulation data.

## 1. Introduction

Basic stochastic methods are intensely used in today's experiments that aim to determine lifetime of unstable nuclei. Most of the already developed methods for calculating mean lifetime of daughter element in the decay chain strongly depend on correct identification of the exact mother-daughter pair in the time sequence data. In case of low activity of the mother isotope (compared to decay constant $\lambda$ of the daughter) it is easy to determine the mother-daughter pair with high probability, but the statistics is low. However, raising the activity will raise the statistics, but decrease the probability for correct pairing of mother-daughter nuclei.

Bernas et. al. (Ref. [1]) analyzed the time correlation between the detection times of a fragment of interest and of a subsequent β particle. In order to obtain the beta decay half-life they formed the distribution of time intervals only between the first β detected after each fragment. They also provided the random distribution by collecting time intervals between the last β occurring before a fragment and the fragment. In this article, we present a technique for determination of mean lifetime of nuclei in decay chain without the need to know the exact relationship between particular decay of the parent and daughter nuclei. This stochastic method is explained in Sec. 2. One of the benefits of this approach is that the increase of activity of mother nuclei is followed by decrease of error of the determined lifetime. Moreover, the method is not activity dependent, meaning that activity may vary during the data acquisition, which may be the case in many real situations. We also pay special attention to determination of errors depending on the activity, for fixed time of measurement and constant activity. The presented concept is checked using an extensive set of Monte Carlo simulation data. The simulation program is custom made and developed by our group. The results of the MC simulation are shown in Sec. 4 and Sec. 5.

## 2. Statistical procedure of formation of time intervals distribution

Let us consider the following decay chain $X \to Y \to Z$. We introduce two parameters that are of importance for our analysis: activity $A(X)$ of the mother nuclide $X$, and the decay constant $\lambda(Y)$ of the daughter nuclide $Y$. Our aim is to determine the mean lifetime of the daughter nuclide, which is the inverse of the decay constant, $\tau = 1/\lambda$. If the radioactive equilibrium is achieved, meaning $A(X) = A(Y)$, we distinguish three cases:

1) If the mean lifetime of the $Y$ nuclide is much longer than the time between two $X$ decays, meaning $A(X) \gg \lambda(Y)$, there is a strong possibility that successive decays of $X$ and $Y$ nuclei do not belong to the same $X \to Y \to Z$ chain.
2) Conversely, if the mean lifetime of the $Y$ nuclide is much shorter than the time between two $X$ decays, meaning $A(X) \ll \lambda(Y)$, there is strong possibility that the decay of a nucleus $X$ is followed by the decay of nucleus $Y$, which is the daughter nucleus of the mentioned nucleus $X$.
3) If $A(X) \sim \lambda$, it is not clear what the relationship between decay of $X$ and the following decay of $Y$ is.

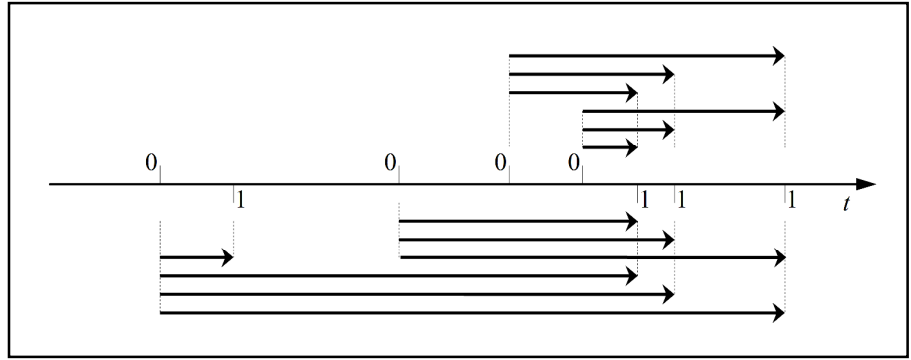

**Fig. 1.** A time sequence with the applied procedure of collecting the time intervals between succeeding start-stop pairs of signals. The start and the stop signals are labeled by 0's (X decay events) and 1's (Y decay events), respectively. The arrows correspond to the collected time intervals.

As far as the second case is concerned, the identification of the right decay is good. It is possible to form a set of time differences between the decays of the mother nuclei and the decays of the corresponding daughter nuclei. The average of these differences is the estimate of the mean lifetime of nucleus $Y$. The downside of this case is that the statistics is low since the activity of nucleus $X$ is low.

We are interested in the first case where the statistics is high but the identification of the right decay is poor. Especially if there is an abundance of nuclei $X$ around the detector, there is a possibility that after the formation of one $Y$ nucleus one or more other $Y$ nuclei may be detected before the first one decay. Because of the simultaneous presence of many $Y$ nuclei it is not possible to determine the order in which the nuclei decay, thus it is not possible to determine the lifetimes of those nuclei (Fig. 1).

We observe an array of signals originating from the decay of $X$ and $Y$ nuclei that are randomly settled on time scale. Start signals arise at the time instants of decay of $X$ nuclei ($X$ event). Also, stop signals arise at instants of decay of $Y$ nuclei ($Y$ event). This array of start and stop signals may be treated in many different ways in order to obtain the decay constant of the daughter radionuclide. One possibility is to join weights for the probability of correct pairing with each pair of start-stop signals formed. In other words, it is possible to calculate the probability for each pair that it is a real coincident pair, as in Ref. [2]. Another solution is to neglect all the cases where a start signal is not followed by a stop signal, but by another start signal; and to keep only sequences with clear start-stop coincident pairs, as it is done in Ref. [3]. Both methods have up and downsides.

Our approach is not to neglect any signal in order to keep high statistics. We pair different start and stop signals, but we do not assign probability of correct pairing for each pair. Instead we build and investigate the time distribution of intervals between a start signal and all the following stop signals, as shown in Fig 1.

Let us denote the following variables:
$A$ – the activity of the parent $X$, $\varepsilon_X$, $\varepsilon_Y$ – the efficiency for detection start signal (decay of $X$) and stop signal (decay of $Y$),

$p_{n(X \to Y)}(t)$ – the probability that in the interval $[0, t]$ exactly $n$ stop signals which originate only from random coincidence between $X$ and $Y$ decay occur,

$p_{n(X \to YC)}(t)$ – the probability that in the interval $[0, t]$ exactly $n$ stop signals occur, in which one true coincidence between $X$ and $Y$ decay is found (there can be only one!),

$p_n(t)$ – the probability that in the interval $[0, t]$ exactly $n$ stop signals occur regardless of the origin,

$p_n^{Coll}(t)$ - the probability that in the interval $[0, t]$ exactly n time intervals are collected - which include detector efficiency $\varepsilon_X$ for detection start signal (decay of $X$) and detector efficiency $\varepsilon_Y$ for detection stop signal (decay of $Y$).

It is clear that:

$$p_n = p_{n(X \to Y)} + p_{n(X \to YC)}$$

The probability to collect $n$ uncorrelated time intervals are given by recursive formula:

$$dp_{(n+1)(X \to Y)}^{Coll}(t) = \varepsilon_X p_{n(X \to Y)}(t')e^{-A(t-t')}e^{-\lambda(t-t')}\varepsilon_Y A dt'$$

$$p_{(n+1)(X \to Y)}^{Coll}(t) = \int_0^t \varepsilon_X p_{n(X \to Y)}(t')e^{-A(t-t')}e^{-\lambda(t-t')}\varepsilon_Y A dt'$$

$$p_{n(X \to Y)}^{Coll}(t) = \varepsilon_X \varepsilon_Y p_{n(X \to Y)}(t), \text{ where } p_{n(X \to Y)}(t) = A\frac{(At)^{n-1}}{(n-1)!}e^{-(A+\lambda)t}$$

The probability to collect $n$ time intervals with one true coincidence:

$$dp_{(n+1)(X \to YC)}^{Coll}(t) = \varepsilon_X p_{n(X \to Y)}(t')e^{-A(t-t')}e^{-\lambda(t-t')}\varepsilon_Y \lambda dt' + \varepsilon_X p_{n(X \to YC)}(t')e^{-A(t-t')}\varepsilon_Y A dt'$$

$$p_{(n+1)(X \to YC)}^{Coll}(t) = \varepsilon_X \varepsilon_Y \int_0^t e^{-A(t-t')}(\lambda p_{n(X \to Y)}(t')e^{-\lambda(t-t')} + A p_{n(X \to YC)}(t'))dt'$$

$$p_{n(X \to YC)}^{Coll}(t) = \varepsilon_X \varepsilon_Y p_{n(X \to YC)}(t), \text{ where }$$

$$p_{n(X \to YC)}(t) = A\frac{(n-1)(At)^{n-2}}{(n-1)!}e^{-(A+\lambda)t}(e^{\lambda t}-1) + \lambda\frac{(At)^{n-1}}{(n-1)!}e^{-(A+\lambda)t}$$

The sum of probability distributions for all possible $n$ (from 1 to infinity) is:

$$p^{Coll}(t) = \sum_{n=1}^{\infty}\left(p^{Coll}_{n(X \to Y)}(t) + p^{Coll}_{n(X \to YC)}(t)\right)$$

$$p^{Coll}(t) = \varepsilon_X \varepsilon_Y \left(A + \lambda e^{-\lambda t}\right) \qquad (1)$$

Probabilities and their sum are shown in Figure 2(b).

In the case when $\varepsilon_X = \varepsilon_Y = 1$ and the possibility to have real $X \to Y$ coincidence could be ignored ($\lambda \ll A$), we get well know equation for combined probability, Ref. [4]:

$$p^{Coll}(t) = A$$

Probabilities and their sum in this case are shown in Figure 2(a).

In the case of constant activity of the parent after normalization the time distribution is given by:

$$f(t) = AT(A + \lambda e^{-\lambda t}) \qquad (2)$$

where $T$ is the total time of measurement.

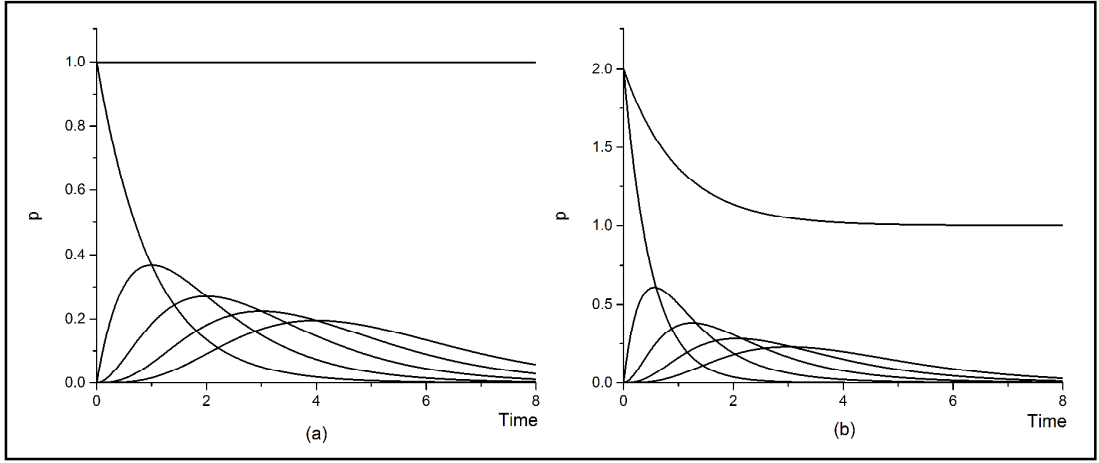

**Fig 2.** Probabilities graphics for the first 5 intervals between start and stop signals and sums of probabilities for all intervals without (a) and with (b) coincidence between start and stop signals

### 3. No distinguish between decay of mother and daughter

In the case that we cannot distinguish decays of the mother and daughter nuclei, besides $p_{(X \to Y)}$ and $p_{(X \to YC)}$ we have to introduce three more probabilities:

$p_{(X \to X)}$ - start and stop signals originate from decay of $X$,

$p_{(Y \to Y)}$ - start and stop signals originate from decay of Y,

$p_{(Y \to X)}$ - start signal is from decay of Y and stop signal is from decay of X.

Using the similar procedure as we have already used, we get:

$$p_{n(X \to X)}^{Coll}(t) = \varepsilon_X^2 p_{n(X \to X)}(t), \quad p_{n(Y \to Y)}^{Coll}(t) = \varepsilon_Y^2 p_{n(Y \to Y)}(t), \quad p_{n(Y \to X)}^{Coll}(t) = \varepsilon_X \varepsilon_Y p_{n(Y \to X)}(t)$$

where

$$p_{n(X \to X)}(t) = p_{n(Y \to Y)}(t) = p_{n(Y \to X)}(t) = A \frac{n-1}{(n-1)!} e^{-At}(At)^{n-2}$$

Summing for all $n$ from 1 to infinity:

$$p_{(X \to X)}^{Coll}(t) = \varepsilon_X^2 A, \quad p_{(Y \to Y)}^{Coll}(t) = \varepsilon_Y^2 A, \quad p_{(Y \to X)}^{Coll}(t) = \varepsilon_X \varepsilon_Y A$$

From formula (1) we already know that for $p_{(X \to Y)}(t)$ and $p_{(X \to YC)}(t)$

$$p^{Coll}(t) = \sum_{n=1}^{\infty} \left( p_{n(X \to Y)}^{Coll}(t) + p_{n(X \to YC)}^{Coll}(t) \right) = \varepsilon_X \varepsilon_Y \left( A + \lambda e^{-\lambda t} \right)$$

Combining these two sums we get:

$$p^{Coll}(t) = \varepsilon_X \varepsilon_Y \left( A + \lambda e^{-\lambda t} \right) + \varepsilon_X^2 A + \varepsilon_X \varepsilon_Y A + \varepsilon_X^2 A = (\varepsilon_X + \varepsilon_Y)^2 A + \varepsilon_X \varepsilon_Y \lambda e^{-\lambda t}$$

It is clear that for $\varepsilon_X = \varepsilon_Y$ (which is the case in most real situation when we cannot distinguish decay of mother and daughter) the stochastic combinatorial background is four time bigger comparing to the situation when we can distinct these two decays. This has direct impact on the error in determination of λ, especially for high $A/\lambda$ ratio.

### 4. Constant activity of the parent

An extensive set of Monte Carlo simulation of decay chain X → Y → Z was done. Time distributions for different $A/\lambda$ are shown in Fig 3. In the case of constant activity A(X), which can be checked and determined from the observed time list, we can use one, two and three parametric fit.

For one parametric fit we can use formula (2), with λ as the fitting parameter.

For two parametric fit the array of start-stop signal pairs is analyzed and fitted using the following function:

$$f(t) = N(A + \lambda e^{-\lambda t}) \qquad (3)$$

where $N$ is normalization factor, λ is the decay constant of the daughter radionuclide and $A$ stands for the constant activity of the parent radionuclide determined from the time list. Equation (3) corresponds to equation (1) of our theoretical concept.

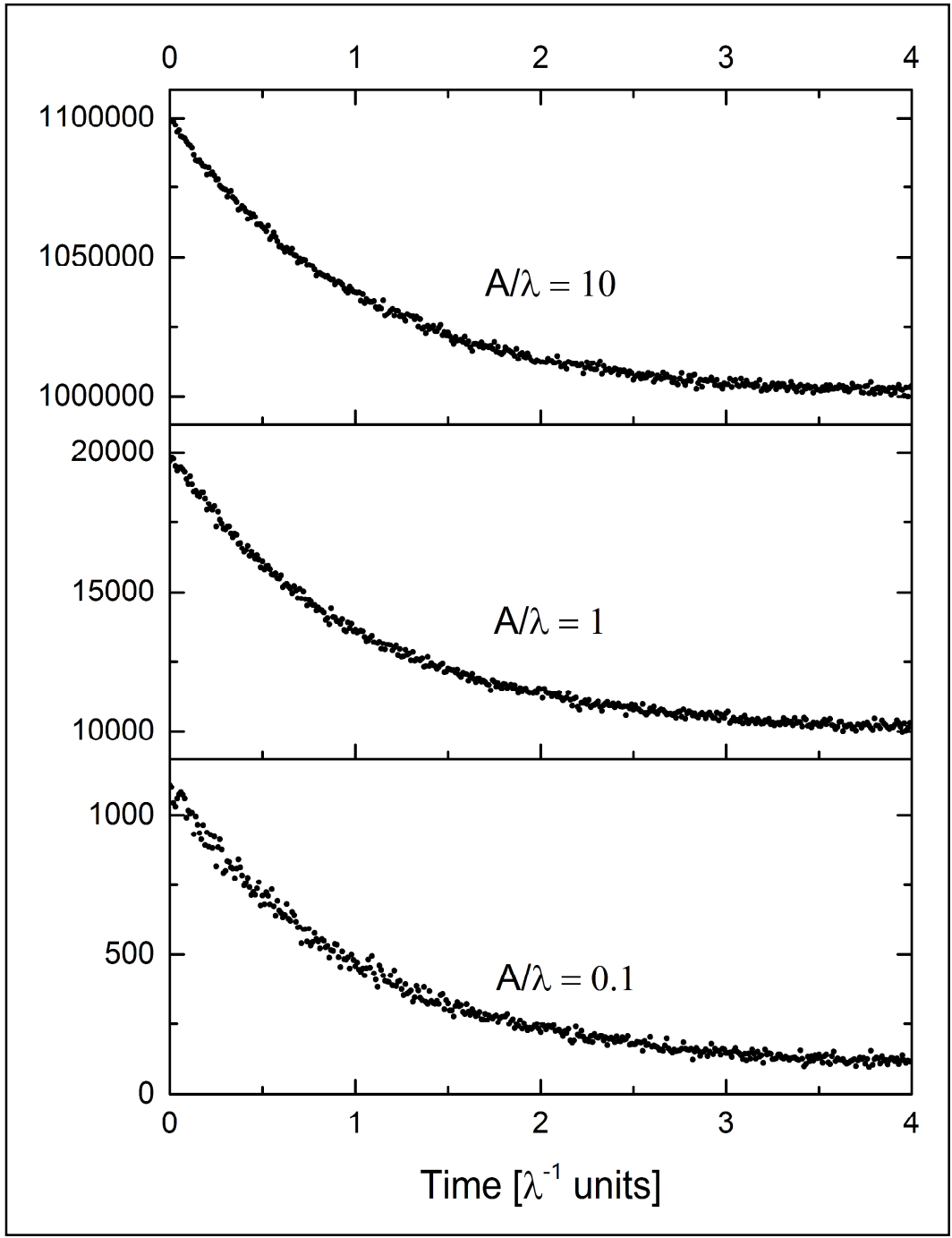

**Fig. 3.** Time distributions for different activities obtained for the fixed simulation time ($10^6$ $\lambda^{-1}$). Time span for fitting procedure was from 0 to $4\lambda^{-1}$.

It is possible to calculate the decay constant without assumption that $A(X)$ = const using fitting function with three parameters:

$$f(t) = C + De^{-\lambda t}, \tag{4}$$

This way the change of the activity of the radioactive source could be taken into account. Detector efficiency also influences the result. For low detecting efficiency, it is necessary to observe the simulation for a longer period of time in order to get the same level of statistical confidence. This means that the simulation time influences the uncertainty of the decay parameters that we try to determine.

The distributions for different (but constant) activities within fixed simulation time ($T_{sim} = 10^6 \lambda^{-1}$) were produced. The results are shown in Fig. 3 and Table 1.

|  | 1 parameter fit | | 2 parameters fit | | 3 parameters fit | |
| --- | --- | --- | --- | --- | --- | --- |
| A/$\lambda$ | $\lambda$ | $\Delta\lambda$ | $\lambda$ | $\Delta\lambda$ | $\lambda$ | $\Delta\lambda$ |
| 0.01 | 1.031 | 0.013 | 1.032 | 0.013 | 1.031 | 0.024 |
| 0.03 | 1.0128 | 0.0078 | 1.0138 | 0.0079 | 1.007 | 0.015 |
| 0.1 | 1.0080 | 0.0050 | 1.0089 | 0.0051 | 1.000 | 0.010 |
| 0.3 | 1.0000 | 0.0035 | 1.0017 | 0.0035 | 1.0092 | 0.0073 |
| 1 | 0.9946 | 0.0027 | 0.9963 | 0.0026 | 1.0009 | 0.0060 |
| 3 | 1.0006 | 0.0023 | 1.0021 | 0.0023 | 1.0011 | 0.0055 |
| 10 | 1.0015 | 0.0021 | 1.0012 | 0.0021 | 0.9989 | 0.0052 |
| 30 | 1.0007 | 0.0023 | 0.9960 | 0.0020 | 0.9990 | 0.0052 |
| 100 | 1.0089 | 0.0030 | 0.9991 | 0.0020 | 0.9985 | 0.0052 |
| 300 | 1.0047 | 0.0023 | 0.9998 | 0.0020 | 1.0007 | 0.0052 |
| 1000 | 0.9917 | 0.0023 | 0.9970 | 0.0020 | 0.9955 | 0.0052 |

**Table 1.** Dependence of the statistical uncertainty of the decay constant on A/$\lambda$ ratio for the fixed simulation time and fit with one, two and three parameters.

The fitting procedure was done for time distributions with one, two and three parameters functions, as in Equ. 2, 3 and 4. Illustration of the results is shown in Table 1. and Fig. 4.

Using possibility to freely set the activity of the mother radionuclide in Monte Carlo data, the study of uncertainty of the decay constant for different activities is done. Fig. 4 shows that the uncertainty gets lower as the activity gets higher. For higher activities this uncertainty is flat. This shows the advantage of this method for cases where the activity is high, exactly what we wanted to analyze.

For the situation that we cannot distinct signature of mother and daughter decays we expect bigger error for calculated $\lambda$ due to the rise of the combinatorial background. The results are shown in Table 2.

|  | Distinctive | | Non distinctive | |
| --- | --- | --- | --- | --- |
| A/$\lambda$ | $\lambda$ | $\Delta\lambda$ | $\lambda$ | $\Delta\lambda$ |
| 0.1 | 1.000 | 0.010 | 0.990 | 0.012 |
| 1 | 1.0009 | 0.0060 | 0.986 | 0.010 |
| 10 | 0.9989 | 0.0052 | 1.019 | 0.010 |
| 30 | 0.9990 | 0.0052 | 0.975 | 0.010 |
| 100 | 0.9985 | 0.0052 | 1.001 | 0.010 |

**Table 2.** Statistical uncertainty of the decay constant on A/$\lambda$ ratio for the fixed simulation time and three parameters fit, in the cases that we can and cannot distinct decays of the mother and daughter.

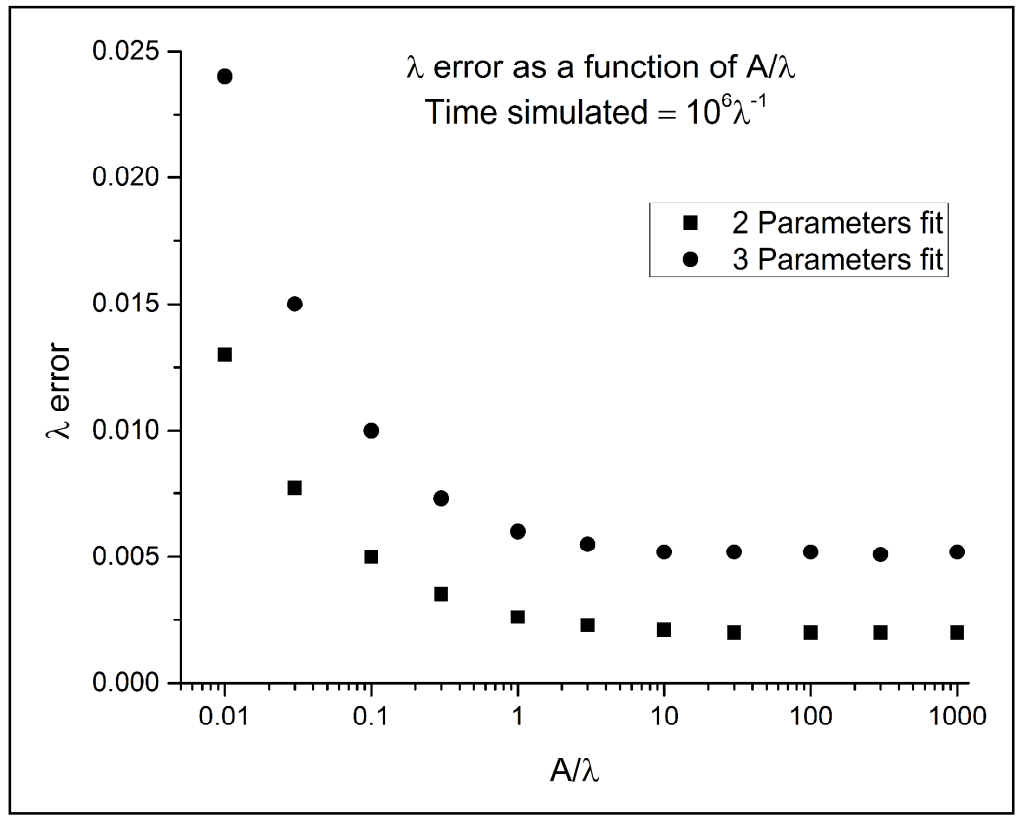

**Fig. 4.** Dependence of the decay constant error of activity to decay constant ratio

### 5. Variable activity of the parent

The case of exponentially decreasing activity $A=A_0e^{-kt}$ was also studied and presented in Fig. 5. Time of simulation was the same, $10^6 \lambda^{-1}$. Starting activity was $10A/\lambda$ and the final was $0.1A/\lambda$. This corresponds to real possible cases where lifetime of the mother nucleus is much lower than the time of measurement.

For two parametric fit, according to Equ. 3, we accept the mean activity during the measurement time for *A*. It is clear from Fig. 5, reduced chi square ($\chi^2$=2.25) and evaluated $\lambda$=0.48 that two parametric fit is not acceptable for this case of changing activity of the parent. However, three parametric fit shows excellent agreement with the predicted value of $\lambda$=1, even in the case of large change of activity of the parent nuclide (100:1)

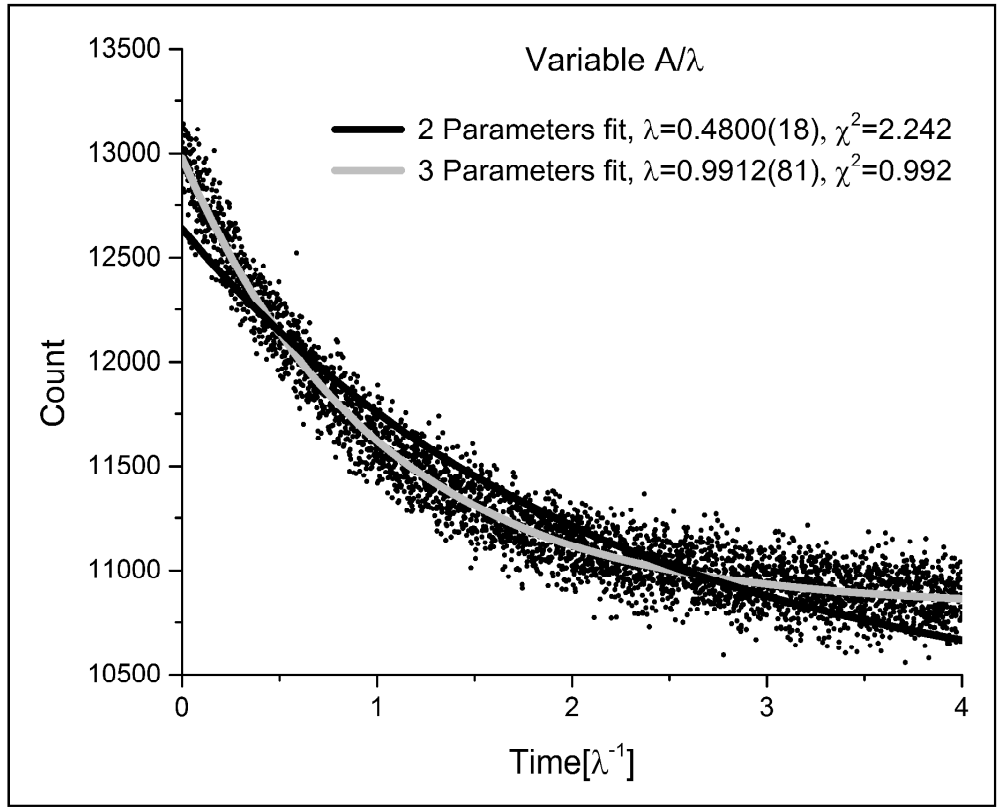

**Fig. 5.** Time distributions for variable mother nucleus activity fitted with two and three parameters functions

## 6. Summary

We have introduced an improved method of determination of the mean lifetime of nuclei in the radioactive decay chain. The method consists of a statistical procedure of formation of time intervals distribution, which we described in detail. We derived the corresponding theoretical time distribution of the intervals between signals generated by simultaneous radioactive decay of many mother-daughter pairs of nuclei. The simple form of the derived distribution of time intervals makes it suitable for determination of the mean lifetime of many short living nuclei which decay in the radioactive chain. A main advantage of this method is that an increase in the activity of the mother nuclei is followed by a decrease in uncertainty of the determined mean lifetime. In addition, the method is not activity dependent, meaning that activity may vary during the data acquisition, which is often the case in many real situations. We tested the method by Monte Carlo simulations which proved all the advantages of the method.

### 7. Acknowledgements

The authors are grateful to the Ministry of Education and Science of the Republic of Serbia for the financial support of the present work, Project No ON171002 and ON171028.